\documentclass[pra,floatfix,twocolumn,showpacs]{revtex4}

\usepackage{amsmath}
\usepackage{amsfonts}
\usepackage{latexsym}
\usepackage{xypic}
\usepackage{epsfig}
\usepackage{graphicx}
\usepackage{dcolumn}

\begin{document}
\title{Controlling spatiotemporal chaos and spiral turbulence in
excitable media: A review}
\author{Sitabhra Sinha}
\email{sitabhra@imsc.res.in}
\author{S. Sridhar}
\affiliation{%
The Institute of Mathematical Sciences, C.I.T. Campus, Taramani,
Chennai - 600 113 India
}%

\begin{abstract}
Excitable media are a generic class of models used to simulate a wide
variety of natural systems including cardiac tissue.
Propagation of excitation waves in this
medium results in the formation of characteristic patterns such as rotating 
spiral waves. Instabilities in these structures may lead to
spatiotemporal chaos through spiral turbulence, which has been linked
to clinically diagnosed conditions such as cardiac fibrillation.
Usual methods for controlling such phenomena involve very large amplitude
perturbations and have several drawbacks. There have been several 
recent attempts to develop low-amplitude control procedures for spatiotemporal 
chaos in excitable media which are reviewed in this paper.
The control schemes have been broadly classified by us into three 
types: (i) global, (ii) non-global spatially-extended and (iii) local, 
depending on the way the control signal is applied, and we discuss the merits 
and drawbacks for each.
\end{abstract}
\maketitle

\section{Introduction}
\label{ssch:sec1}
Excitable media 
denotes a class of systems that share a set of features
which make their dynamical behavior qualitatively similar. These features
include (i) the existence of two characteristic dynamical states, comprising
a stable {\em resting state} and a metastable {\em excited state}, 
(ii) a {\it threshold} value associated with one of the dynamical
variables characterising the system, on exceeding which, the system switches
from the resting state to the excited state, and (iii) a {\em recovery period}
following an excitation, during which the response of the system to a
supra-threshold stimulus is diminished, if not completely 
absent~\cite{ss:Keener98}. 
Natural systems which exhibit such features include, in biology, cells 
such as neurons, cardiac myocytes 
and pancreatic beta cells, all of which
are vital to the function of a complex living organism. Other examples
of dynamical phenomena associated with excitable media include 
cAMP waves observed 
during aggregation of slime mold, calcium waves observed 
in Xenopus oocytes, muscle contractions during childbirth in uterine tissue, 
chemical waves observed in the Belusov-Zhabotinsky 
reaction and concentration
patterns in CO-oxidation reaction on Pt(110) surface.
Excitation
in such systems is observed as the characteristic {\em action potential},
where a variable associated with the system (e.g., membrane potential, 
in the case of
biological cells) increases very fast from its resting value to the peak
value corresponding to the excited state, followed by a slower process
during which it gradually returns to
the resting state.

The simplest model system capable of exhibiting all these features is the
generic Fitzhugh-Nagumo set of coupled differential equations:
\begin{equation}
de/dt = e(1-e)(e-b) - g, ~~dg/dt = \epsilon (ke - g),
\label{FHNeq}
\end{equation}
which, having only two variables, is obviously incapable of exhibiting chaos.
However, when several such sets are coupled together diffusively to simulate
a spatially extended media (e.g., a piece of biological tissue made
up of a large number of cells), the resulting high-dimensional dynamical
system can display chaotic behavior. The genesis of this {\em spatiotemporal
chaos} 
lies in the distinct property of interacting waves in excitable media,
which mutually annihilate on colliding. This is a result of the fact 
that an excitation wavefront is followed by a region whose
cells are all in the recovery period,
and which, therefore, cannot be stimulated by another excitation wavefront,
as for example when two waves cross each other~\cite{ss:note1}. 
Interaction between such waves result in the
creation of spatial patterns, referred to variously as {\em reentrant 
excitations} (in 1D), {\em vortices} or {\em spiral waves}
(in 2D) and {\em scroll waves} (in 3D), which form when an 
excitation wavefront is broken as the wave propagates across partially
recovered tissue or encounters an inexcitable obstacle~\cite{ss:Jalife99}. 
The free ends
of the wavefront gradually curl around to form
spiral waves. Once formed, such waves become self-sustained sources of
high-frequency excitation in the medium, and usually can only be terminated
through external intervention. 
The existence of nonlinear properties of wave propagation in several excitable
media
can lead to
complex non-chaotic spatiotemporal rhythms.
Thus, spiral waves are associated
with periodic as well as quasiperiodic patterns of temporal activity.

\begin{figure}[tb]
\includegraphics[width=.95\linewidth]{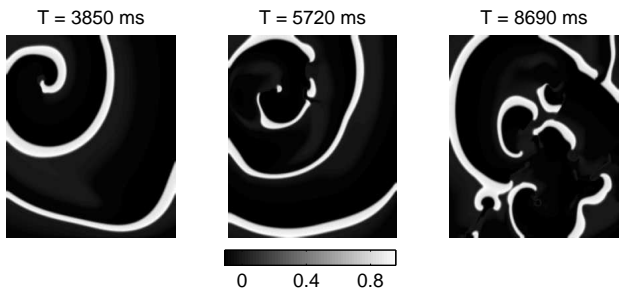} 
\caption{Onset of spatiotemporal chaos in the 2-dimensional Panfilov model
of linear dimension $L = 256$. The initial condition is a broken plane wave
that is allowed to curl around into a spiral wave (left). Meandering 
of the spiral focus causes wavebreaks to occur (centre) that eventually
result in spiral turbulence, with multiple independent sources of
high-frequency excitation (right).}
\label{ss:fig0}
\end{figure}

However, in this paper, we shall
not be discussing the many schemes proposed to terminate single spiral waves,
but instead, focus on the control of spatiotemporally chaotic patterns seen in
excitable media (in 2 or 3 dimensions), that occur when under certain 
conditions, spiral or scroll waves become unstable and break up.
Various
mechanisms of such breakup have been identified~\cite{ss:note2},
including meandering of the
spiral focus. If the meandering is sufficiently high, the spiral wave can 
collide with itself and break up spontaneously, resulting in
the creation of multiple smaller spirals (Fig.~\ref{ss:fig0}). 
The process continues until 
the spatial extent of the system is spanned by several coexisting spiral 
waves that activate different regions without any
degree of coherence. This state of {\em spiral turbulence} marks the
onset of spatiotemporal chaos, as indicated by the Lyapunov spectrum and
Kaplan-Yorke dimension \cite{ss:Pandit02}.

Controlling spatiotemporal chaos in excitable media has certain special
features. Unlike other chaotic systems, here the response to a control signal
is not proportional to the signal strength because of the existence
of a threshold. As a result, an excitable system shows discontinuous response
to control. For instance, regions that have not yet recovered from a previous
excitation or where the
control signal is below the threshold, will not be affected by the
control stimulus at all. Also, the focus of control in excitable 
media is to eliminate all activity rather than to stabilize
unstable periodic behavior. This is because the problem of chaos 
termination has great practical importance in the clinical context, as the 
spatiotemporally chaotic state has been associated
with the cardiac problem of ventricular fibrillation (VF).
VF involves
incoherent activation of the heart that results in the cessation of pumping
of blood, and is fatal within minutes in the absence of external intervention.
At present, the only effective treatment is electrical defibrillation,
which involves applying very strong electrical shocks across the heart
muscles, either externally using a defibrillator or internally through 
implanted devices. The principle of operation for such devices is to
overwhelm the natural cardiac dynamics, so as to drive all the
different regions of the heart to rest simultaneously, at which time the 
cardiac pacemaker can take
over once again. Although the exact mechanism by which this is achieved 
is still not
completely understood, the danger of using such large amplitude control
(involving $\sim kV$ externally and $\sim 100 V$ internally) is that,
not only is it excruciatingly painful to the patient, but by causing damage
to portions of cardiac tissue which subsequently result in scars, 
it can potentially increase the likelihood of future arrhythmias.
(i.e., abnormalities in the heart's natural rhythm).
Therefore, devising a low-power control method for 
spatiotemporal chaos
in excitable media promises a safer treatment for people at risk from 
potentially fatal cardiac arrhythmias.

In this paper, we have discussed most of the recent control methods
that have been proposed for terminating spatiotemporal chaos in excitable
media~\cite{ss:note3}.
These methods are also
often applicable to the related class of systems known as oscillatory media, 
described by complex Landau-Ginzburg equation \cite{ss:Aranson02}, 
which also exhibit spiral waves
and spatiotemporal chaos through spiral breakup.
We have broadly classified all control schemes into three types, depending on
the nature of application of the control signal. If every region of the media
is subjected to the signal (which, in general, can differ from region to region)
it is termed as {\em global control}; 
on the other hand, if the control
signal is applied only at a small, localised region from which its effects
spread throughout the media, this is called {\em local control}. 
Between
these two extremes lie control schemes where perturbations are
applied simultaneously to
a number of spatially distant regions. We have termed these methods as
{\em non-global, spatially extended control}. While global control may be
the easiest to understand, involving as it does the principle of synchronizing
the activity of all regions, it is also the most difficult to implement in 
any practical situation. On the other hand, local control
will be the easiest to implement (requiring a single control point)
but hardest to achieve. 

In the next section we describe a few of the more commonly used models
for studying control of spatiotemporal chaos in excitable media. Section 3
discusses proposed methods of global control, while Section 4 discusses other
spatially extended schemes. The next section deals with local control methods,
and we conclude with a brief section containing general discussions about
chaos control and its implications.
\section{Models of Spatiotemporal Chaos in Excitable Media}
\label{ssch:sec2}
\begin{figure}[tb]
\includegraphics[width=.95\linewidth,clip]{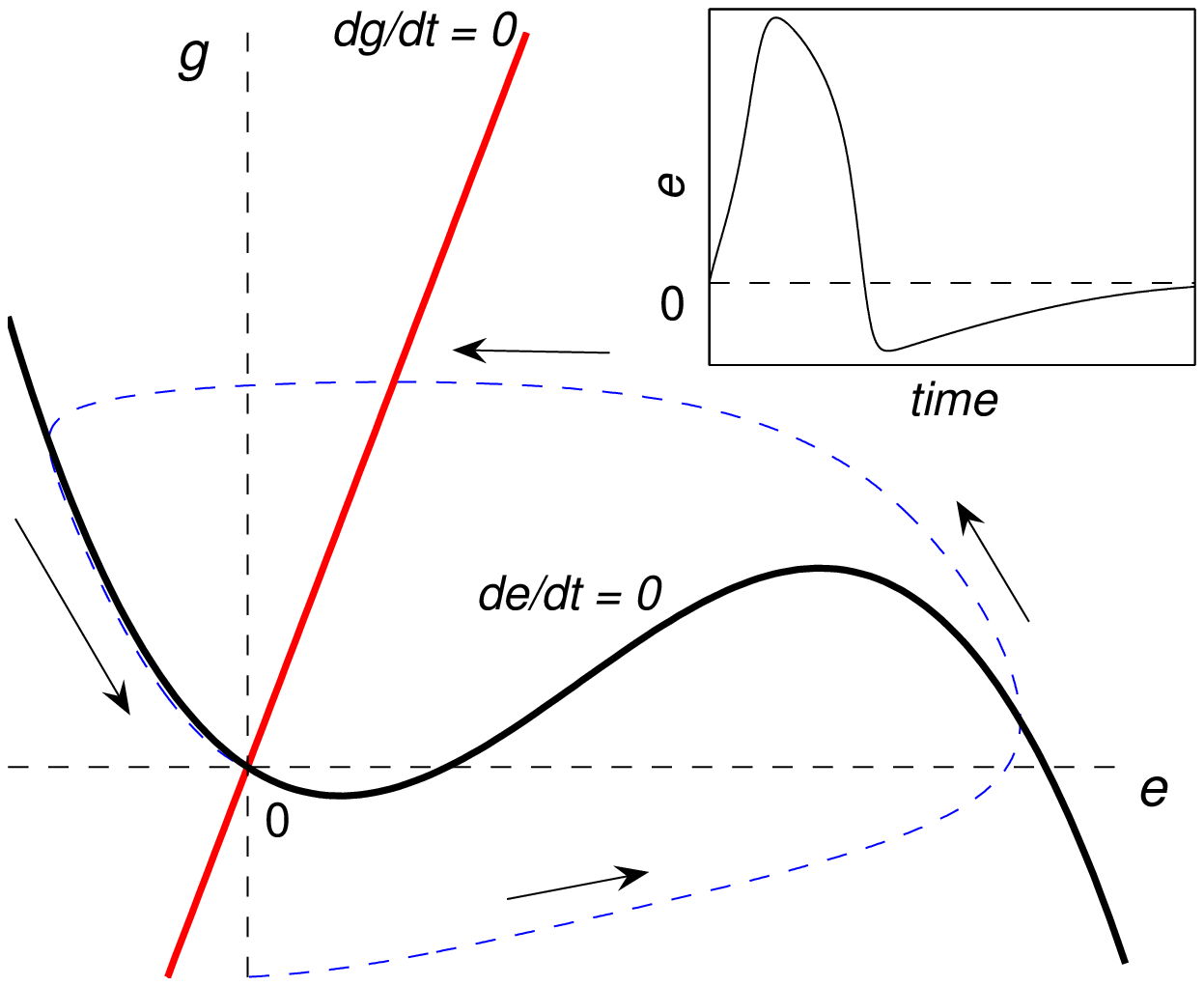} 
\includegraphics[width=.95\linewidth,clip]{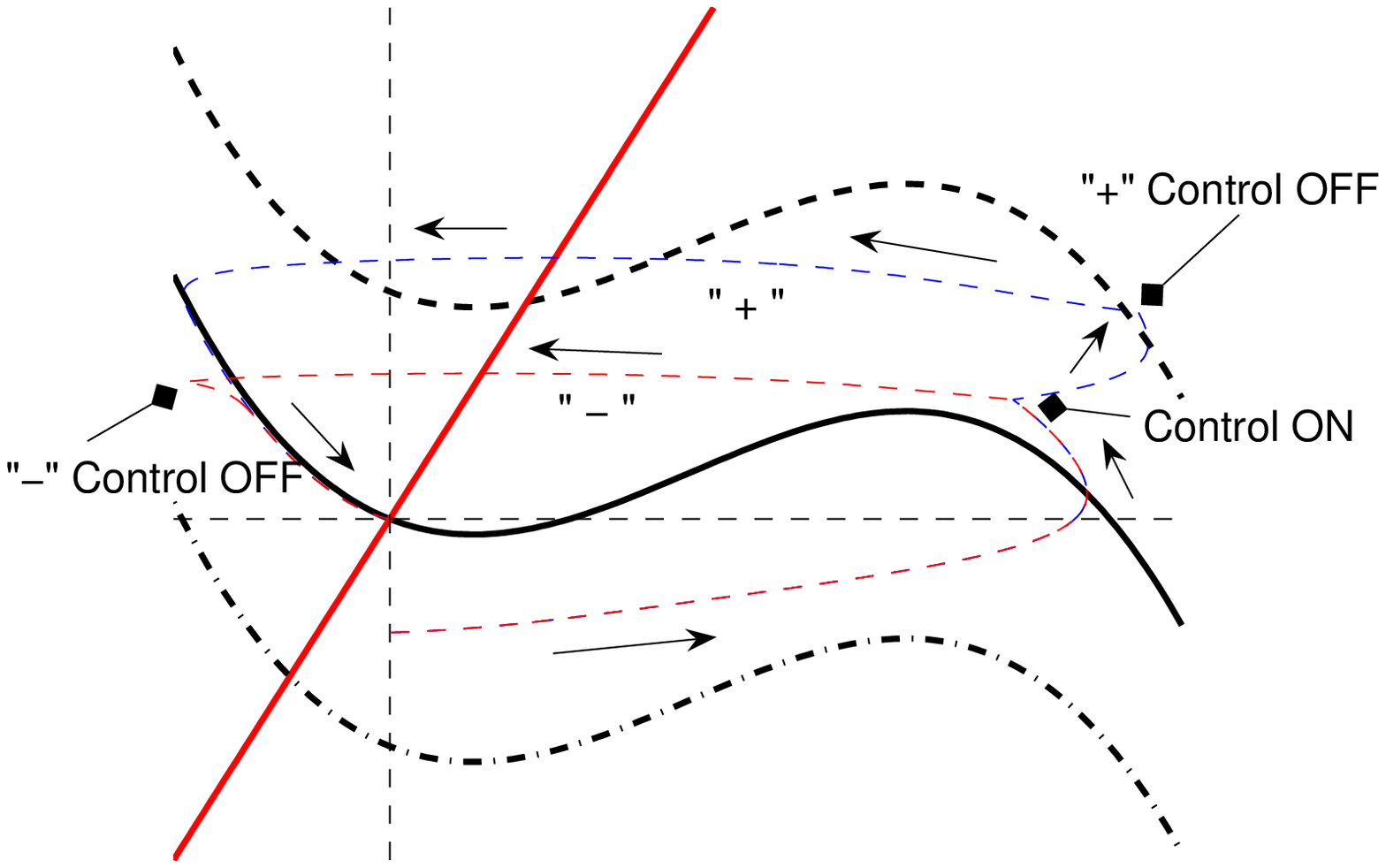}
\caption{(top) Dynamics in the phase-space of the Fitzhugh-Nagumo model, with
the resulting time evolution of the action potential shown in the inset. 
The resting state corresponds to $e = 0, g = 0$.
(Bottom) The result
of applying a positive (``+'') or negative (``$-$'') additive perturbation 
of the same duration to the $e$ variable: ``+'' control decreases the 
threshold and makes excitation more likely, while ``$-$'' control 
decreases the duration
of the action potential and allows the system to recover faster. 
For the duration of the control signal, the $e$-nullcline shifts upward 
(downward) for positive (negative) perturbation as indicated
by the dashed (dash-dotted) curve.}
\label{ss:fig1a}
\end{figure}

The generic Fitzhugh-Nagumo model for excitable media (Eq. \ref{FHNeq}) 
exhibits a structure that is common to most models used in the papers
discussed here. Typically, the dynamics is described by a fast variable,
$e({\bf x},t)$,
and a slow variable, $g({\bf x},t)$, the ratio of 
timescales being given by $\epsilon$.
The resulting phase space behavior is shown in Fig.~\ref{ss:fig1a} (left).
For biological cells, the fast variable is often associated with the
transmembrane potential, while the slow (recovery) variable represents 
an effective
membrane conductance that replaces the complexity of several different
types of ion channels.
For the spatially extended system, the fast variable
of neighboring cells are coupled diffusively. 
There are several models belonging to this general class of excitable media
which display breakup of spiral waves (in 2D) and scroll waves (in 3D), 
including the one 
proposed by Panfilov~\cite{ss:Panfilov93,ss:Panfilov98}
\begin{equation}
{\partial e}/{\partial t} = {\nabla}^2
e - f(e) - g, ~~
{\partial g}/{\partial t} = {\epsilon}(e,g) (ke - g).
\label{Panfeq}
\end{equation}
Here, $f(e)$ is the function specifying
the initiation of the action potential
and is piecewise linear:
$f(e)= C_1 e$, for $e<e_1$, $f(e) = -C_2 e + a$,
for $e_1 \leq e \leq e_2$, and $f(e) = C_3 (e - 1)$, for $e > e_2$.
The physically appropriate parameters given
in Ref.~\cite{ss:Panfilov98} are $e_1 = 0.0026$, $e_2 = 0.837$, $C_1 = 20$,
$C_2 = 3$, $C_3 = 15$, $a = 0.06$ and $k = 3$. The function
$\epsilon (e,g)$ determines the time scale for the dynamics of the
recovery variable: $\epsilon (e,g) = \epsilon_1$ for
$e < e_2$, $\epsilon (e,g) = \epsilon_2$ for $e > e_2$, and
$\epsilon (e,g) = \epsilon_3$ for $e < e_1$ and $g < g_1$ with
$g_1 = 1.8\/$, $\epsilon_1 = 1/75\/$, $\epsilon_2 = 1.0\/$, and
$\epsilon_3=0.3$ 

Simpler variants that also display spiral wave breakup in 2D include
(i) the Barkley model~\cite{ss:Barkley90}:
\begin{equation}
{\partial e}/{\partial t}  =  {\nabla}^2 e +
{\epsilon}^{-1} e (1 - e)(e - \frac{g+b}{a}),~~
{\partial g}/{\partial t}  =  e - g,
\label{Barkleyeq}
\end{equation}
the appropriate parameter values being given in Ref.~\cite{ss:Alonso03},
and (ii) the
B\"{a}r-Eiswirth model~\cite{ss:Bar93}, which differs 
from (\ref{Barkleyeq}) only in having 
${\partial g}/{\partial t} = f(e) - g$,
the functional form of $f(e)$ and parameter values being 
as in Ref.~\cite{ss:Woltering02}.
The Aliev-Panfilov model \cite{ss:Aliev96} is a modified form of the 
Panfilov model, that takes 
into account nonlinear effects such as the
dependence of the
action potential duration on the distance of the wavefront to the
preceding waveback. It has been 
used for control in Refs. \cite{ss:Sakaguchi03,ss:Sakaguchi05}. 


All the preceding models tend to disregard several complex features
of actual biological cells, e.g., the different types of ion channels
that allow passage of electrically charged ions across the cellular
membrane. 
There exists a class of models inspired by the Hodgkin-Huxley formulation
describing action potential generation in the squid giant axon, which explicitly
takes such details into account. While the simple models described above do
reproduce generic features of several excitable media seen in nature,
the more realistic models describe many properties of specific systems,
e.g., cardiac tissue. The general form of such models are described
by a partial differential equation for the transmembrane potential $V$,
\begin{equation}
\frac{\partial V}{\partial t}+\frac{I_{ion}}{C} = D\nabla^2V,
\end{equation}
where $C$ is the membrane capacitance density and
$D$ is the diffusion constant, which, if the medium is isotropic, is a scalar.
$I_{ion}$ is the instantaneous total ionic-current density, and different
realistic models essentially differ in its formulation. For example, in the
Luo-Rudy I model \cite{ss:Luo91} of guinea pig ventricular cells, 
$I_{ion}$ is assumed to be composed of six different ionic current densities,
which are themselves determined by several time-dependent 
ion-channel gating variables whose time-evolution is governed by
ordinary differential equations of the form:
\begin{equation}
\frac{d\xi}{dt} = \frac{\xi_{\infty}-\xi}{\tau_{\xi}}.
\end{equation}
Here, $\xi_{\infty}=\alpha_{\xi}/(\alpha_{\xi}+\beta_{\xi})$ is
the steady state value of $\xi$ and
$\tau_{\xi}=\frac{1}{\alpha_{\xi}+\beta_{\xi}}$ is its time constant.
The voltage-dependent rate constants, $\alpha_{\xi}$ and $\beta_{\xi}$,
are complicated functions of $V$ obtained by fitting experimental data.

\section{Global Control}
\label{ssch:sec3}
The first attempt at controlling chaotic activity in excitable media
dates back almost to the beginning of the field of chaos control itself, when
proportional perturbation feedback (PPF) control was used to stabilize
cardiac arrhythmia in a piece of tissue from rabbit 
heart \cite{ss:Garfinkel92}. This method applied small electrical stimuli,
at intervals calculated using a feedback protocol, to stabilize an unstable
periodic rhythm. Unlike in the original proposal for controlling
chaos \cite{ss:Ott90}, where the location of the stable manifold
of the desired unstable periodic orbit (UPO) was moved using small 
perturbations, 
in the PPF method it is the state of the system that is moved onto the 
stable manifold. However, it has been later pointed out that PPF does
not necessarily require the existence of UPOs (and, by extension, deterministic
chaos)
and can be used even in systems with stochastic dynamics 
\cite{ss:Christini95}. Later, PPF method was used to control atrial 
fibrillation in human heart \cite{ss:Ditto00}. However, the effectiveness
of such control in suppressing spatiotemporal chaos, when applied only at a 
local region, has been questioned, especially
as other experimental attempts in feedback control have not been able to 
terminate fibrillation
by applying control stimuli at a single spatial location \cite{ss:Gauthier02}.

\begin{figure}[tbp]
\includegraphics[width=.95\linewidth]{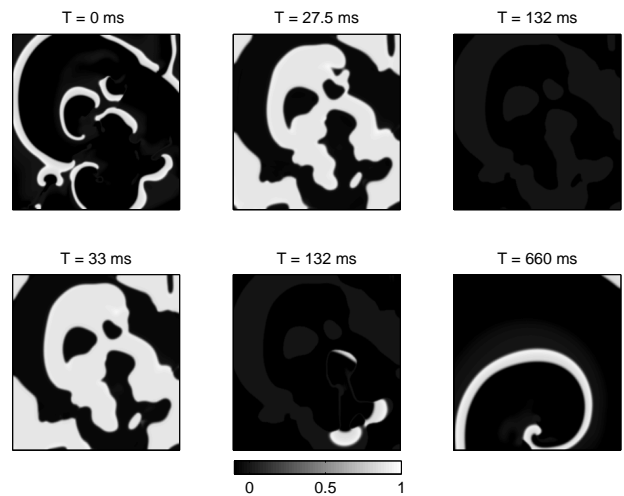}
\caption{Global control of the 2-dimensional Panfilov model with $L = 256$
starting from a spatiotemporally chaotic state (top left). Pseudo-gray-scale
plots of excitability $e$ show the result of applying a pulse of amplitude
$A = 0.833$ between $t =$ 11 ms and 27.5 ms (top centre) that eventually
leads to elimination of all activity (top right). Applying the pulse
between $t =$ 11 ms and 33 ms (bottom left) results in some regions becoming
active again after the control pulse ends (bottom centre) eventually 
reinitiating spiral waves (bottom right).}
\label{ss:fig1}
\end{figure}

More successful, at least in numerical simulations, have been schemes where
control stimuli is applied throughout the system. Such global control
schemes either apply small perturbations to the dynamical variables ($e$ or $g$)
or one of the parameters (usually the excitation threshold). 
The general scheme involves introducing an external control signal $A$
into the model equations, e.g., in the Panfilov model [Eq.~(\ref{Panfeq})]:
\begin{equation}
{\partial e}/{\partial t} = {\nabla}^2 e - f(e) - g + A,
\end{equation}
for a control duration $\tau$. 
If $A$ is a small, positive perturbation, added to the fast variable,
the result is an effective reduction of the threshold (Fig.~\ref{ss:fig1a},
bottom), thereby making simultaneous excitation of different regions more
likely.
In general, $A$ can be periodic,
consisting of a sequence of pulses. Fig.~\ref{ss:fig1} shows the results of
applying a pulse of fixed amplitude but varying durations. While in general,
increasing the amplitude, or the duration, increases the likelihood
of suppressing spatiotemporal chaos,
it is not a simple, monotonic relationship. Depending on the initial
state at which the control signal is applied, even a high amplitude (or
long duration) control signal may not be able to uniformly excite all regions
simultaneously. 
As a result, when the control signal is withdrawn, the inhomogeneous
activation results in a few regions becoming active again and restarting
the spatiotemporal chaotic behavior.

Most global control schemes are variations or modifications of the above
scheme. 
Osipov and Collins \cite{ss:Osipov99} have shown that a low-amplitude
signal used to change 
the value of the slow variable at the front and back of an excitation
wave can result in different wavefront and waveback velocities which
destabilizes the traveling wave, eventually terminating all activity,
and, hence, spatiotemporal chaos.
Gray~\cite{ss:Gray02} has investigated the termination of spiral wave breakup
by using both short and long-duration pulses applied on the fast variable, 
in 2D and 3D systems. This study concluded that while short duration pulses
affected only the fast variable, long duration pulses affected both fast and
slow variables and that the latter is more efficient, i.e., uses less power, in
terminating spatiotemporal chaos. The external control signal can also
be periodic [$A = F sin (\omega t)$], in which case the critical amplitude
$F_c$ required for terminating activity has been found to be a function of the 
signal frequency $\omega$ \cite{ss:Sakaguchi03}. 

\begin{figure*}[tb]
\includegraphics[width=.24\textwidth]{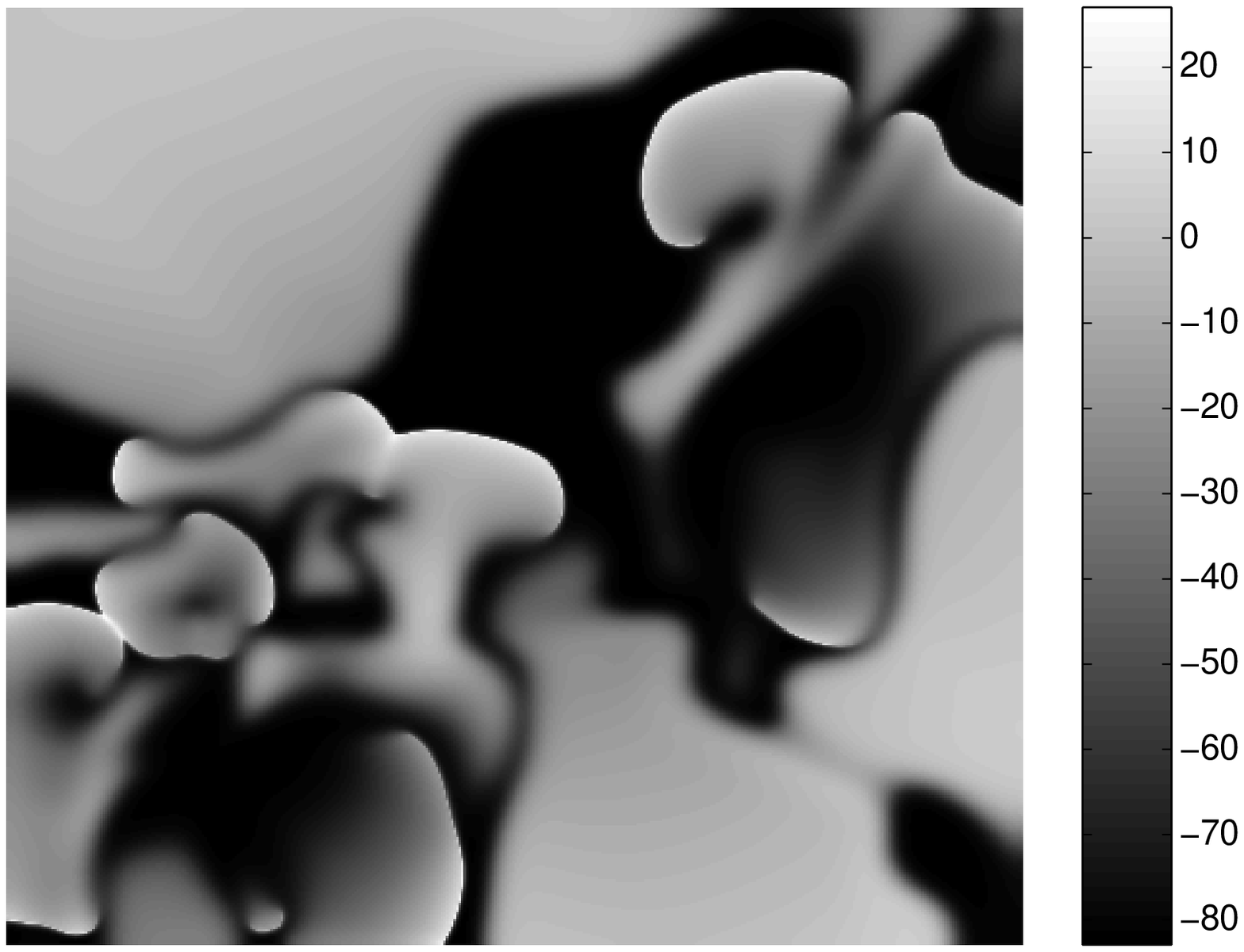}
\includegraphics[width=.24\textwidth]{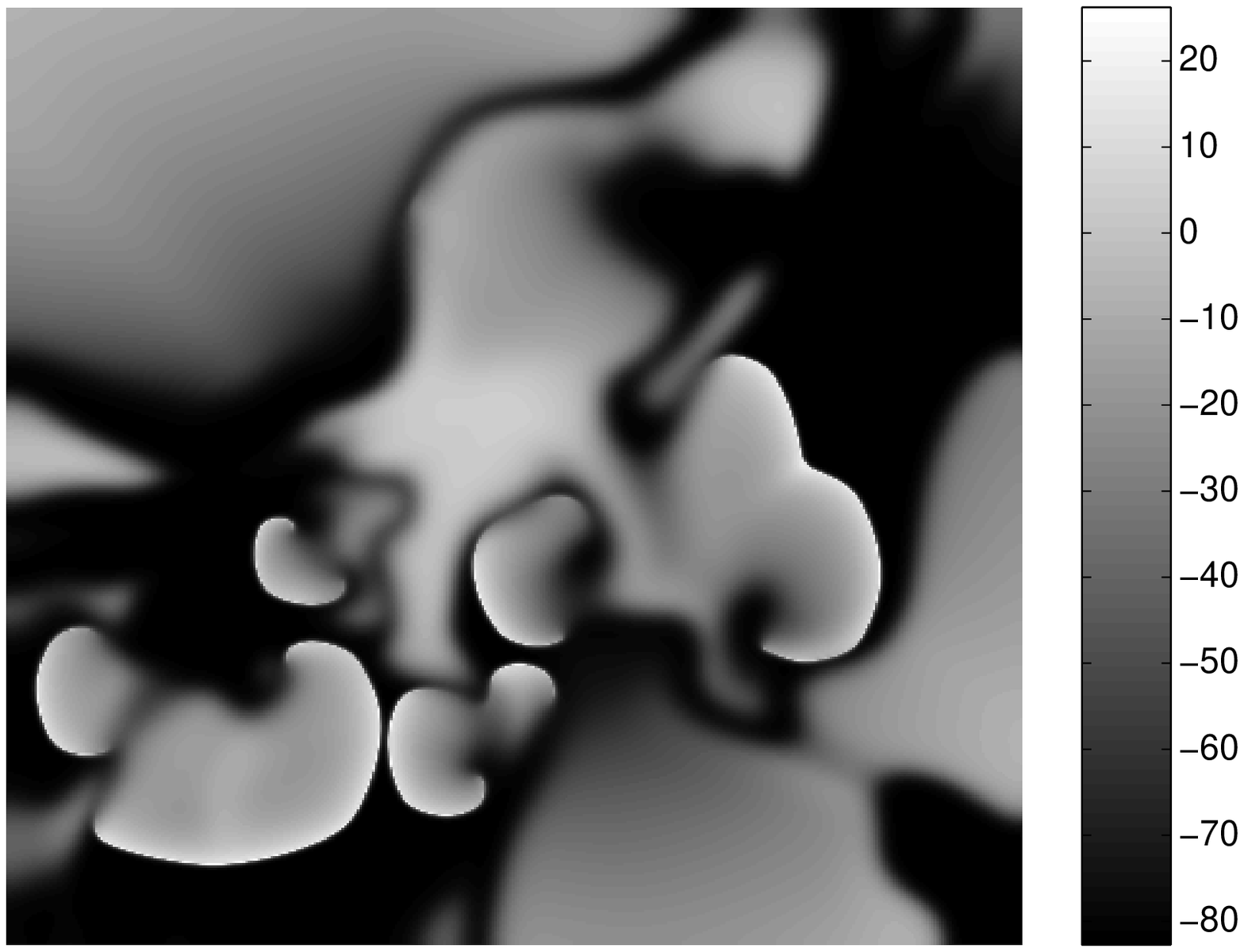}
\includegraphics[width=.24\textwidth]{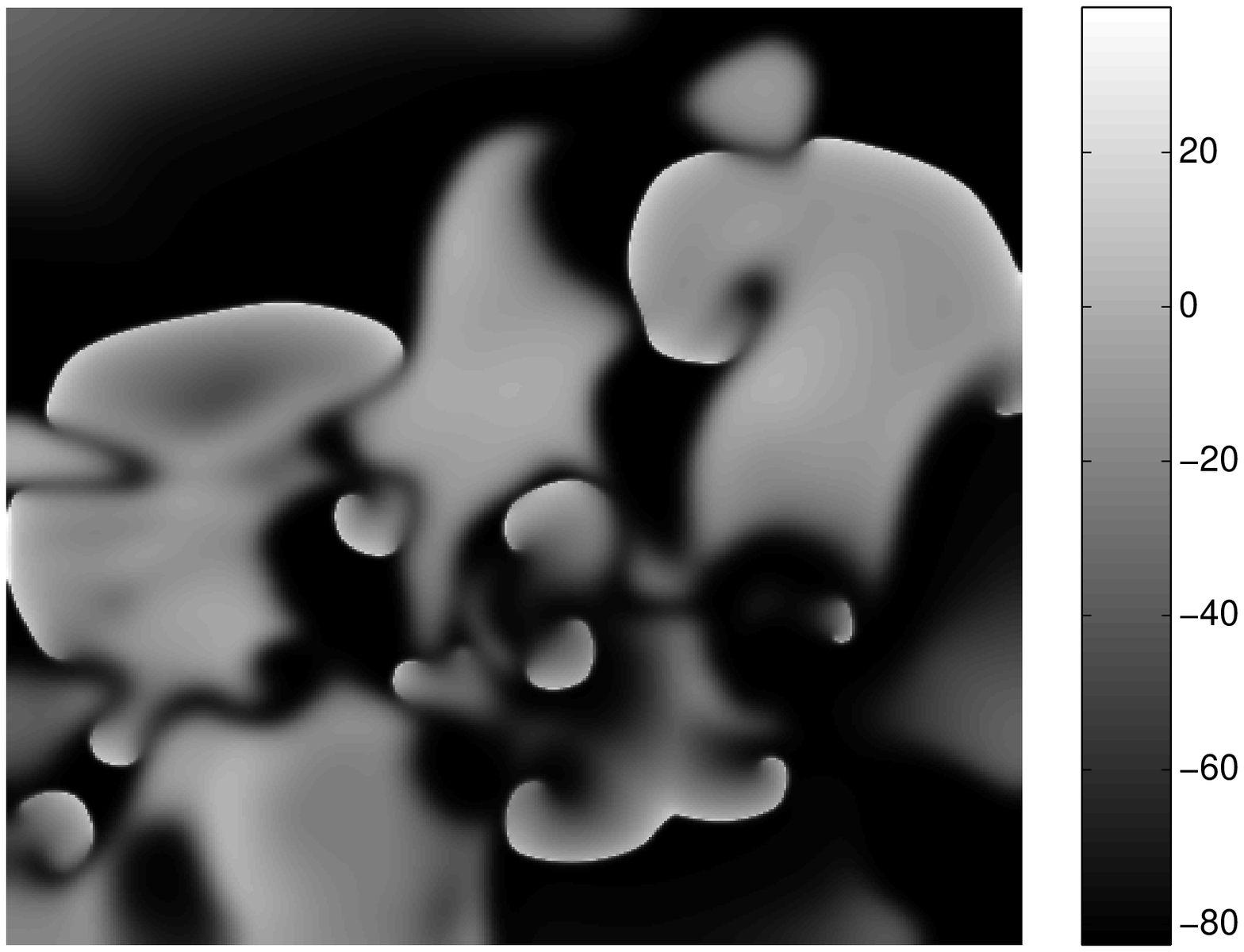}
\includegraphics[width=.24\textwidth]{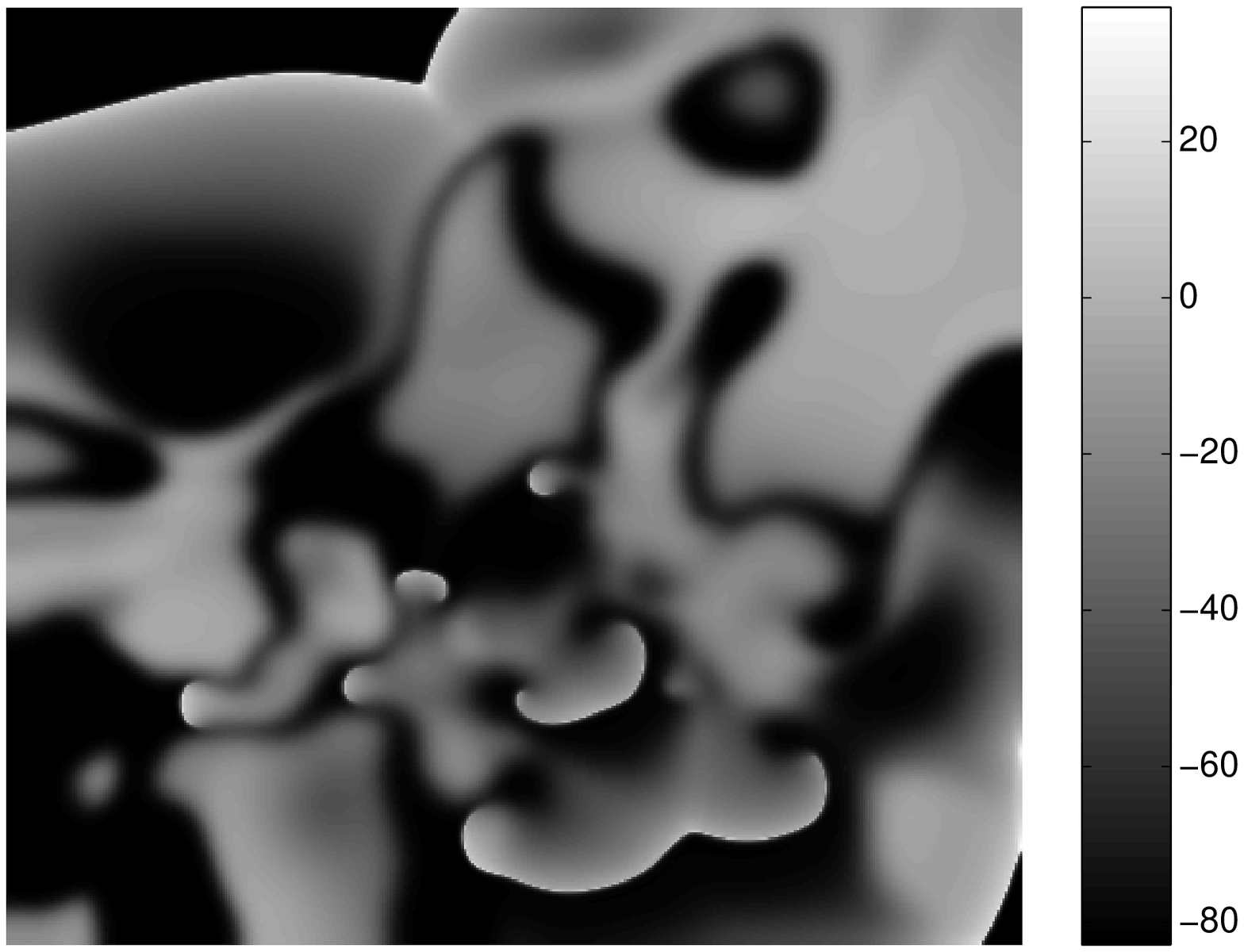}
\includegraphics[width=.24\textwidth]{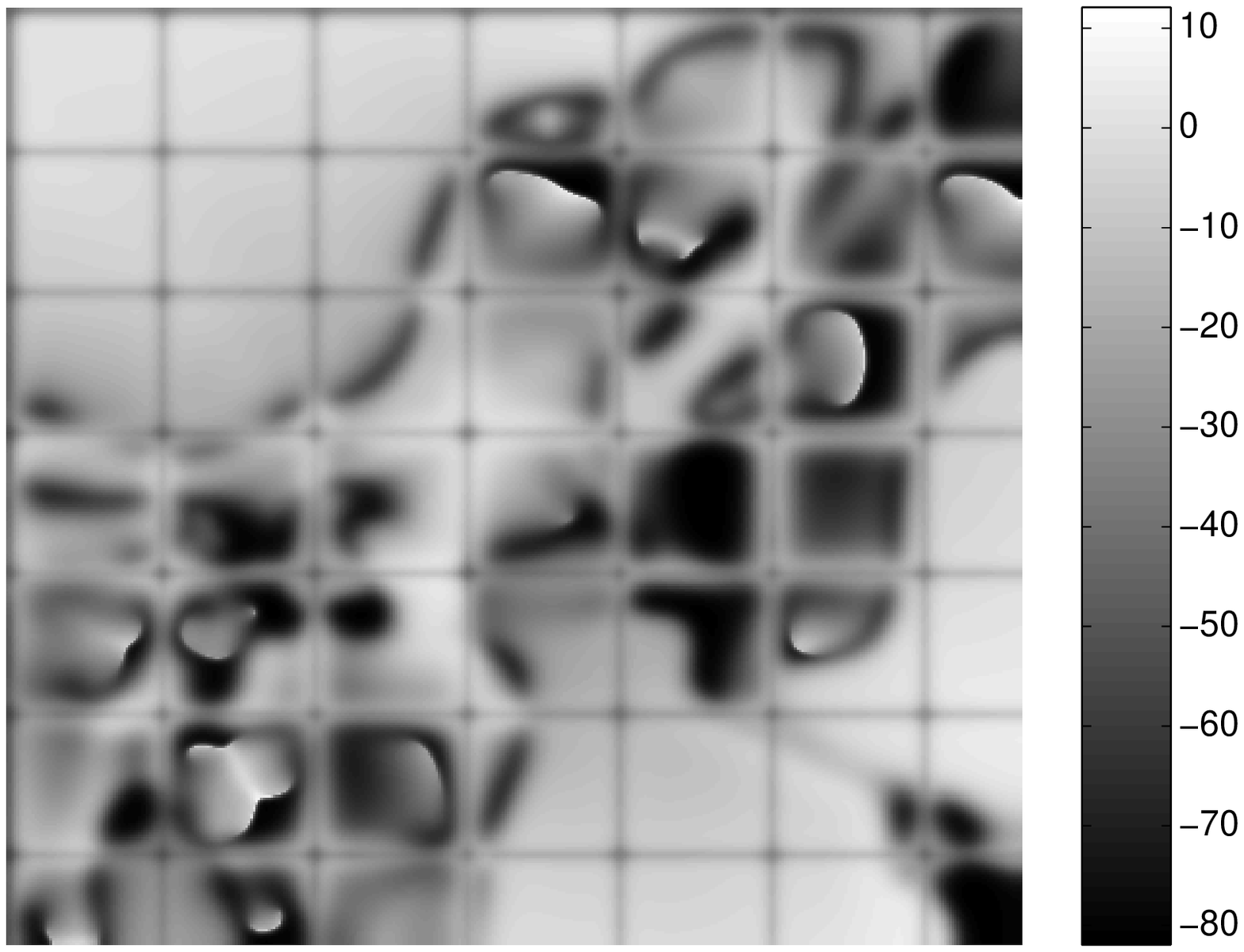}
\includegraphics[width=.24\textwidth]{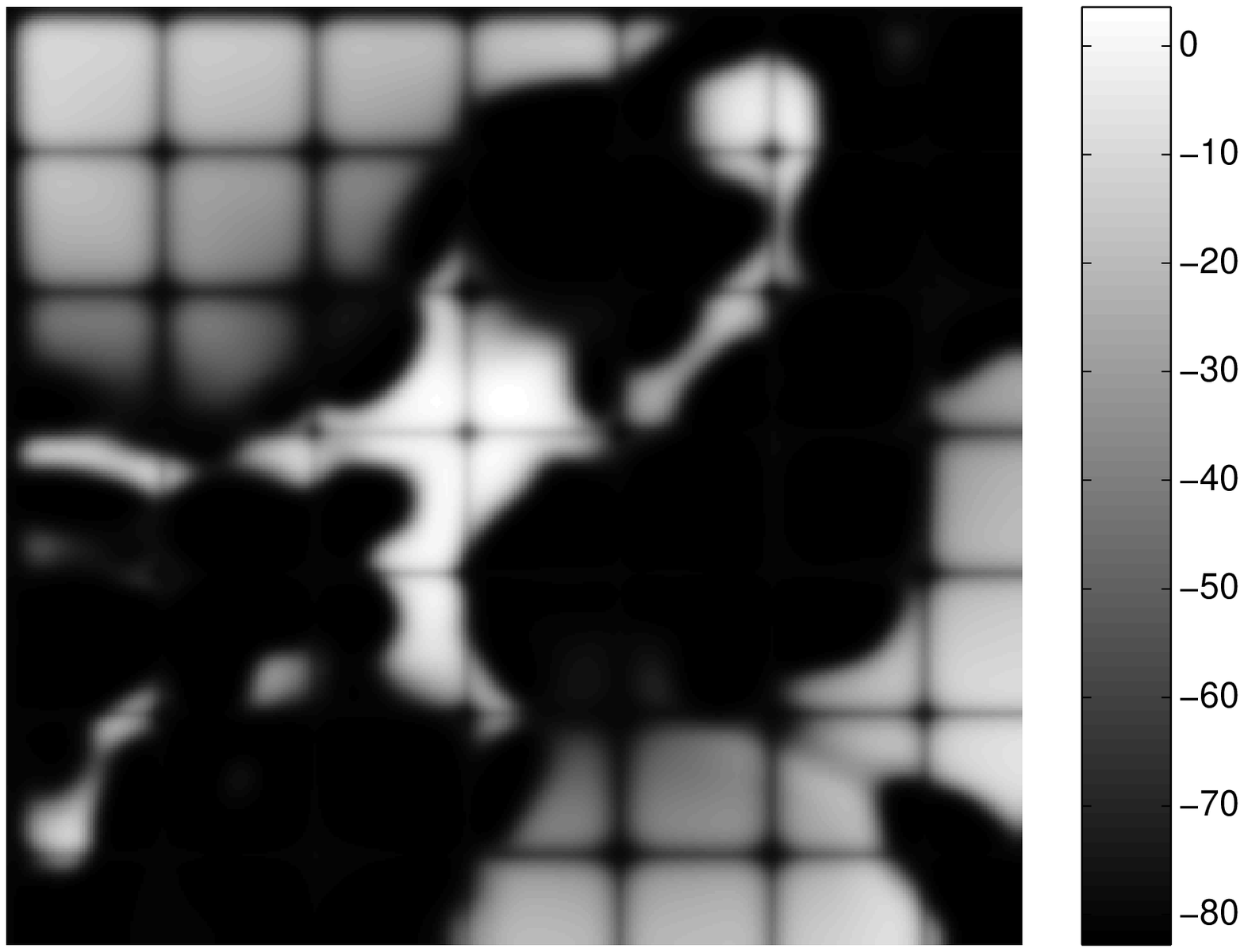}
\includegraphics[width=.24\textwidth]{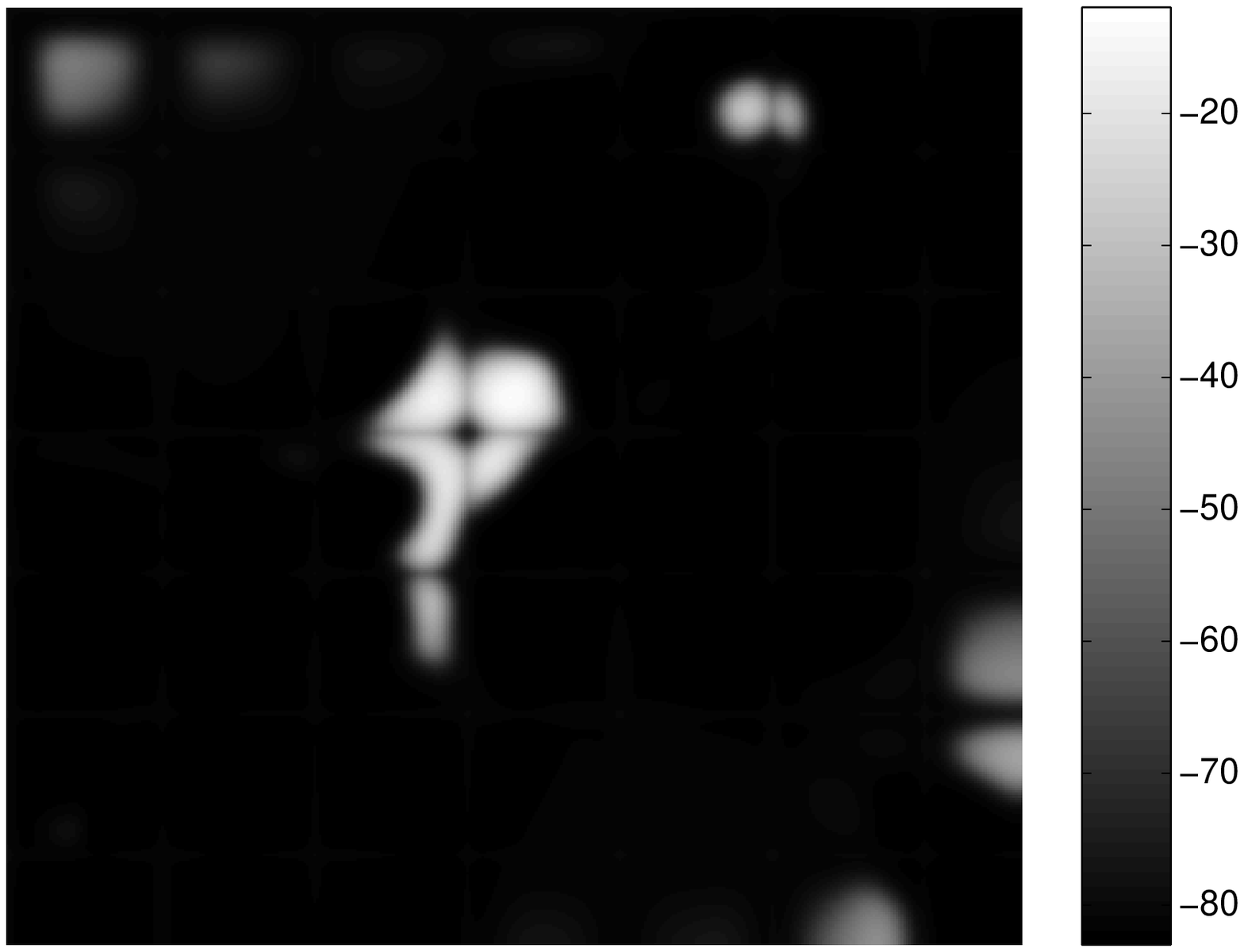}
\includegraphics[width=.24\textwidth]{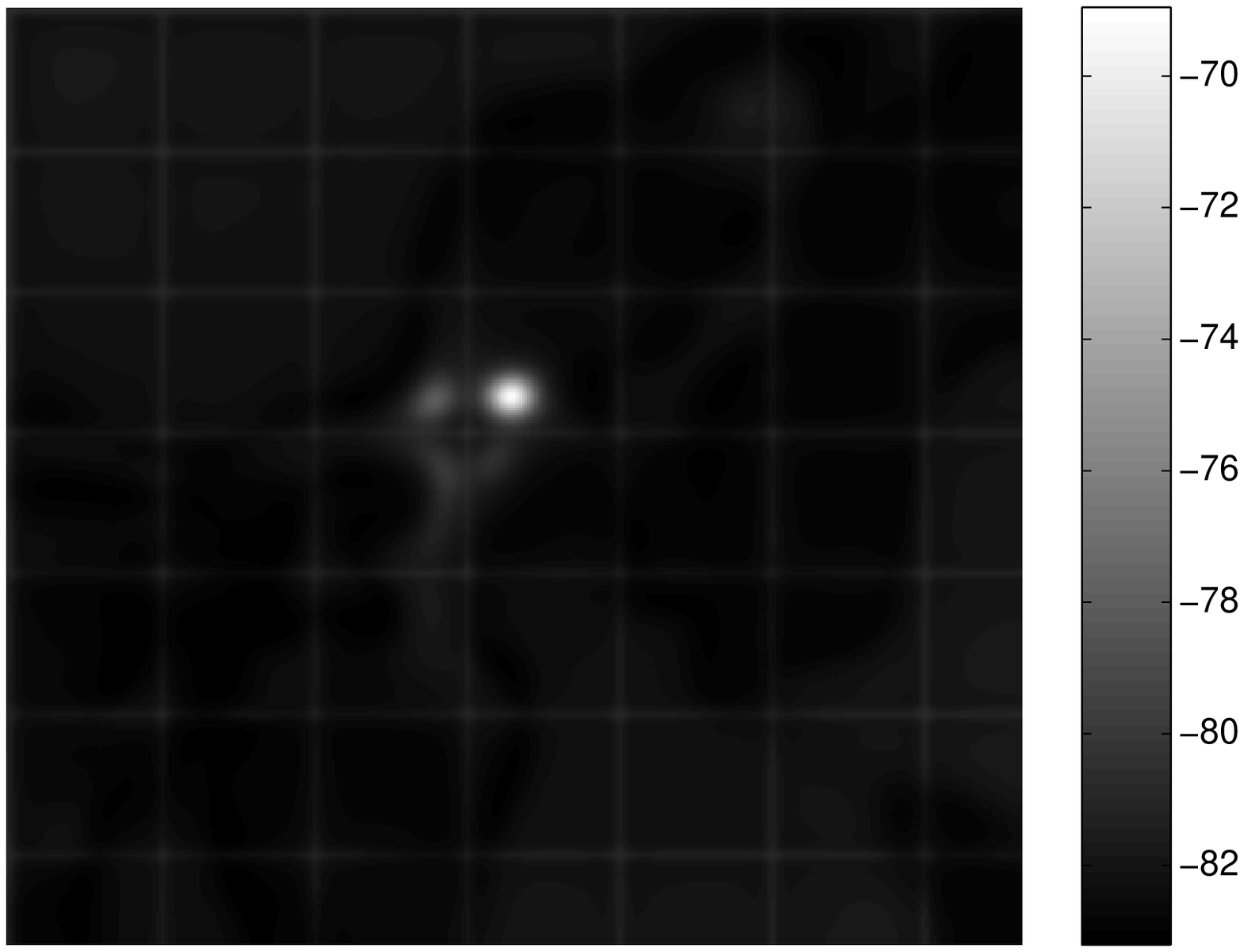}
\caption{Spatiotemporal chaos (top row) and its control (bottom row) in 
the 2-dimensional Luo-Rudy I model with $L = 90$ mm.
Pseudo-gray-scale
plots of the transmembrane potential $V$ show the evolution 
of spiral turbulence at
times $T$ = 30 ms, 90 ms, 150 ms and 210 ms. Control is achieved by 
applying an external current density $I = 150 \mu A/cm^2$ for 
$\tau$ = 2.5 msec over a square
mesh with each block of linear dimension $L/K = 1.35$ cm.
Within 210 msec of applying control, most
of the simulation domain has reached a transmembrane potential
close to the resting state value; moreover, the entire
domain is much below the excitation threshold.
The corresponding uncontrolled case shows spatiotemporal chaos across
the entire domain.}
\label{ss:fig3}
\end{figure*}
Other schemes have proposed applying perturbations to the parameter controlling
the excitation threshold, $b$. Applying a control pulse on this parameter
($b= b_f$, during duration of control pulse;$b=b_0$, otherwise) 
has been shown to cause splitting of an excitation wave into a pair 
of forward and backward moving waves \cite{ss:Woltering02}. 
Splitting of a spiral wave causes the two newly created spirals to 
annihilate each other on collision. For
a spatiotemporally chaotic state, a sequence of such pulses may cause
termination of all excitation, there being an optimal time interval between 
pulses that results in fastest control. Another control scheme that also
applies perturbation to the threshold parameter is the uniform periodic
forcing method
suggested by Alonso {\em et al} \cite{ss:Alonso03,ss:Alonso06} for controlling
scroll wave turbulence in three-dimensional excitable media. Such turbulence
results from negative tension between scroll wave filaments,
i.e., the 
line joining the phase singularities 
about which the scroll wave rotates.
In this control method, 
the threshold is varied in periodic manner [$b = b_0 + b_f cos ( \omega t)$]
and the result depends on the relation between the control frequency $\omega$
and the spiral rotation frequency. If the former is higher than the latter, 
sufficiently strong forcing is seen to eliminate turbulence; otherwise, 
turbulence suppression is not achieved. The mechanism underlying termination
has been suggested to be the effective increase of filament tension due to rapid
forcing, such that, 
the originally negative tension between scroll wave filaments
is changed to positive tension. This results in expanding scroll wave filaments
to instead shrink and collapse, eliminating spatiotemporal chaotic activity.
In a variant method, the threshold parameter has been perturbed by spatially
uncorrelated Gaussian noise, rather than a periodic signal, which also results
in suppression of scroll wave turbulence \cite{ss:Alonso04}.

As already mentioned, global control, although easy to understand, is difficult
to achieve in experimental systems. A few cases in which such control
could be implemented include the case of eliminating spiral wave patterns
in populations of the Dictyostelium amoebae by spraying a fine mist of
cAMP onto the agar surface over which the amoebae cells grow \cite{ss:Lee01}.
Another experimental system where 
global control has been implemented is the photosensitive Belusov-Zhabotinsky
reaction, where a light pulse shining over the entire system is used as a 
control signal \cite{ss:Munuzuri97}. Indeed, conventional defibrillation
can be thought of as a kind of global control, where a large amplitude
control signal is used to synchronize the phase of activity at all
points by either exciting a previously unexcited region (advancing the phase) 
or slowing the recovery of an already excited region (delaying the phase)
\cite{ss:Gray05}.

\section{Non-Global Spatially Extended Control}
\label{ssch:sec4}
The control methods discussed so far apply control signal to all points
in the system. As the chaotic activity is spatially extended, one may
naively expect that any control scheme also has to be global. However, we 
will now discuss some schemes that, while being spatially extended, do
not require the application of control stimuli at all points of the system.
\subsection{Applying control over a mesh}
The control method of Sinha {\em et al} \cite{ss:Sinha01} involving
suprathreshold stimulation along a grid of points 
is based on the
observation that spatiotemporal chaos in excitable media is a long-lived
transient that lasts long enough to establish a non-equilibrium
statistical steady state displaying spiral turbulence. 
The lifetime
of this transient, ${\tau}_L$, increases rapidly
with linear size of the system, $L$, e.g., increasing from 850 ms to
3200 ms as $L$ increases from 100 to 128 in the two-dimensional 
Panfilov model. This accords with the
well-known observation that small mammals do not get
life-threatening VF spontaneously whereas large
mammals do \cite{ss:Winfree87} and has been experimentally verified
by trying to initiate VF in swine ventricular tissue while 
gradually reducing its mass \cite{ss:Kim97}. 
A related observation is that non-conducting boundaries
tend to absorb spiral excitations, which results in spiral waves not lasting
for appreciable periods in small systems.

The essential idea of the control scheme is that a domain can be divided 
into electrically disconnected regions by creating boundaries composed of 
recovering cells between them. These boundaries can be created
by triggering excitation across a thin strip. For two-dimensional
media, the simulation domain (of size $L \times L$) is divided into
$K^2$ smaller blocks by a network of lines with
the block size ($L/K \times L/K$) small enough so that spiral waves cannot form.
For control in a 3D system,
the mesh is used only on one of the
faces of the simulation box.
Control is achieved by applying a suprathreshold stimulation via the mesh
for a duration $\tau$. A network of excited and subsequently recovering
cells then divides the simulation domain into square blocks whose length 
in each direction is fixed at a constant value $L/K$ for the duration of 
control.
The network effectively simulates non-conducting boundary conditions (for the
block bounded by the mesh) for the duration of its recovery period,
in so far as it absorbs spirals formed
inside this block.
Note that $\tau$ need not be large at all because the
individual blocks into which the mesh divides the
system (of linear size $L/K$) are so small that they do not
sustain long spatiotemporally chaotic transients. Nor does $K$,
which is related to the mesh density, have to be very large
since the transient lifetime, $\tau_L$, decreases rapidly with
decreasing $L$.
The method has been applied to multiple excitable models, including
the Panfilov and Luo-Rudy models 
(Fig.~\ref{ss:fig3}). 

An alternative method \cite{ss:Sakaguchi05} for controlling spiral turbulence 
that also uses
a grid of control points has been demonstrated for the Aliev-Panfilov model.
Two layers of excitable media are considered,
where the first layer represents the two-dimensional excitable media 
exhibiting spatiotemporal chaos that is to be controlled, and the second
layer is a grid structure also made up of excitable media.
The two layers are coupled using the fast variable but with asymmetric 
coupling constants, 
with excitation pulses travelling $\sqrt{D}$ times faster in the second
layer compared to the first. As the second layer consists only of grid lines,
it is incapable of exhibiting chaotic behavior in the uncoupled state.
If the coupling from the second layer to the first layer is sufficiently 
stronger than the other way round, the stable dynamics
of the second layer (manifested as a single rotating spiral) overcomes the
spiral chaos in the first layer, and drives it to an 
ordered state characterized by mutually synchronized spiral waves. 

\subsection{Applying control over an array of points}
An alternative method of spatially extended control is to apply 
perturbations at a series of points arranged in a regular array.
Rappel {\em et al} \cite{ss:Rappel99} had proposed using such an arrangement
for applying a time-delayed feedback 
control scheme. However, this scheme
does not control spatiotemporal chaos 
and is outside the scope of this review.

\begin{figure}[tbp]
\includegraphics[width=0.99\linewidth,clip]{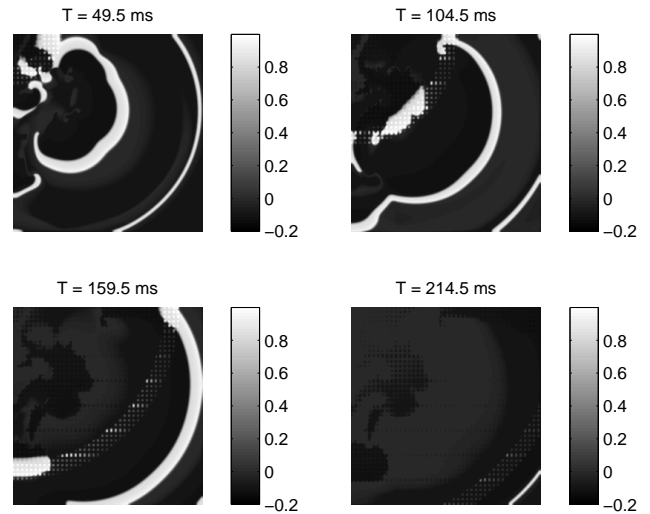}
\caption{Control of the 2-dimensional Panfilov model ($L = 256$) using
an array of control points with spacing $d = 6$ and strength of control
stimulus $A = 2.5$. Stimulation is started at the top left corner ($T = 0$ ms)
and lasts at each control point, as the wave reaches that point, for 17.9 ms.
By 200 ms, the spatiotemporal chaos has disappeared.}
\label{ss:fig4}
\end{figure}
More recently, the authors \cite{ss:Sridhar06}
have used an array of control points to terminate
spatiotemporal chaos in the Panfilov model. Fig.~\ref{ss:fig4} shows the
result of applying a spatially non-uniform control scheme, which simulates
an excitation wave traveling over the system, with the same wavefront 
velocity as in the actual medium. The control points are placed distance $d$
apart along a regular array. At certain times, the control points at one
corner of the system is stimulated, followed by the successive stimulation of
the neighboring control points, such that a wave of stimulation is seen
to move radially away from the site of original stimulation. This process
is repeated after suitable intervals. Note that, simulating a traveling 
wave 
using the array is found to be more effective at controlling spatiotemporal
chaos than the simultaneous activation of all control points.
Using a traveling
wave allows the control signal to engage all high-frequency sources of 
excitation in the spiral turbulence regime, ultimately resulting in complete 
elimination of chaos. If, however, the control had only been applied locally
the resulting wave could only have interacted with neighboring spiral waves
and the effects of such control would not have been felt throughout the system. 
The efficacy of the control scheme depends upon the spacing between the
control points, as well as the number of simulated traveling waves.
Traveling waves have previously been used in Ref.~\cite{ss:Pengye03}
to control spatiotemporal chaos, although in the global control context
with a spatiotemporally periodic signal being applied continuously for
a certain duration, over the entire system.

\section{Local Control of Spatiotemporal Chaos}
\label{ssch:sec5}
\begin{figure*}[htb]
\includegraphics[width=.31\textwidth]{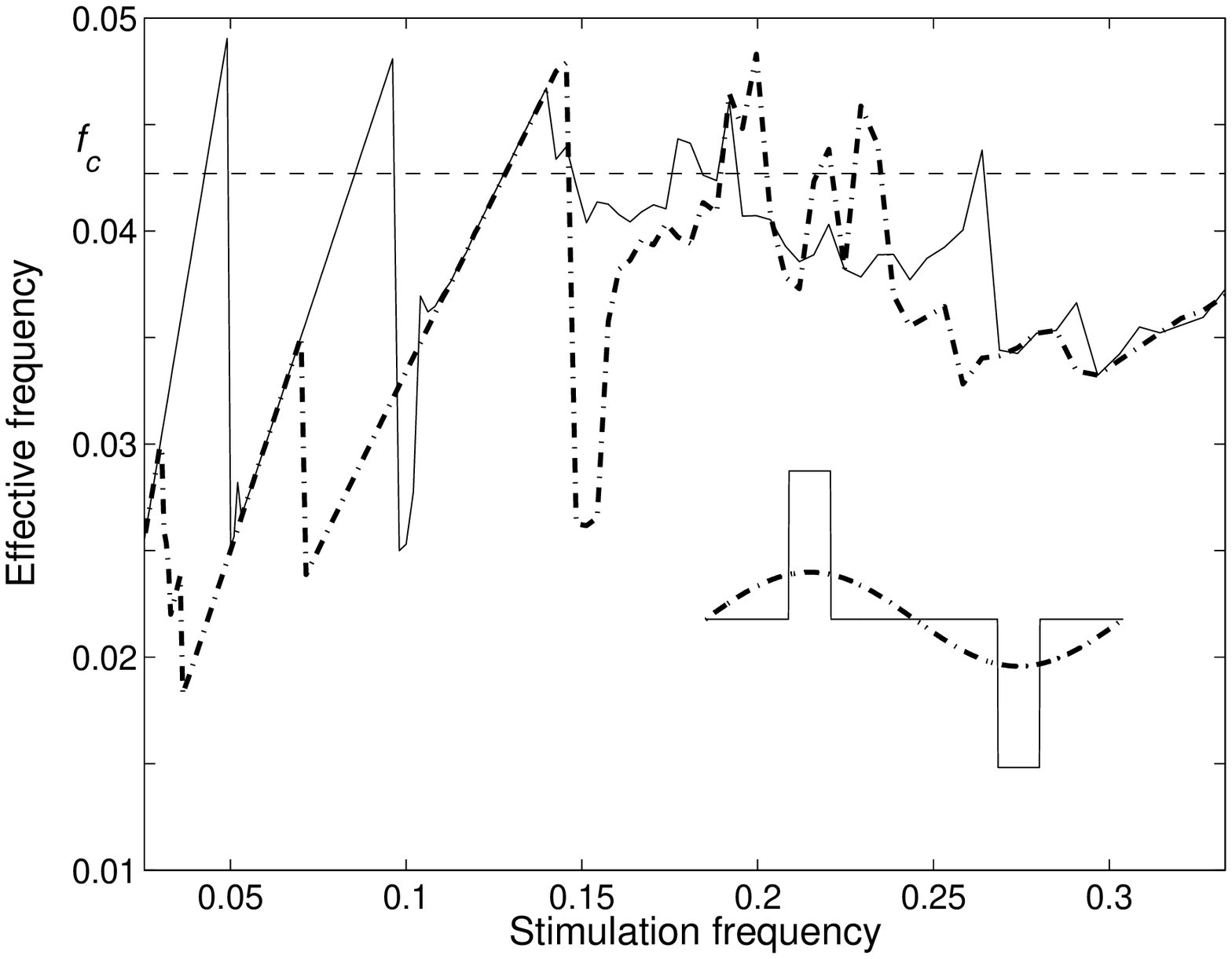}~~
\includegraphics[width=.24\textwidth]{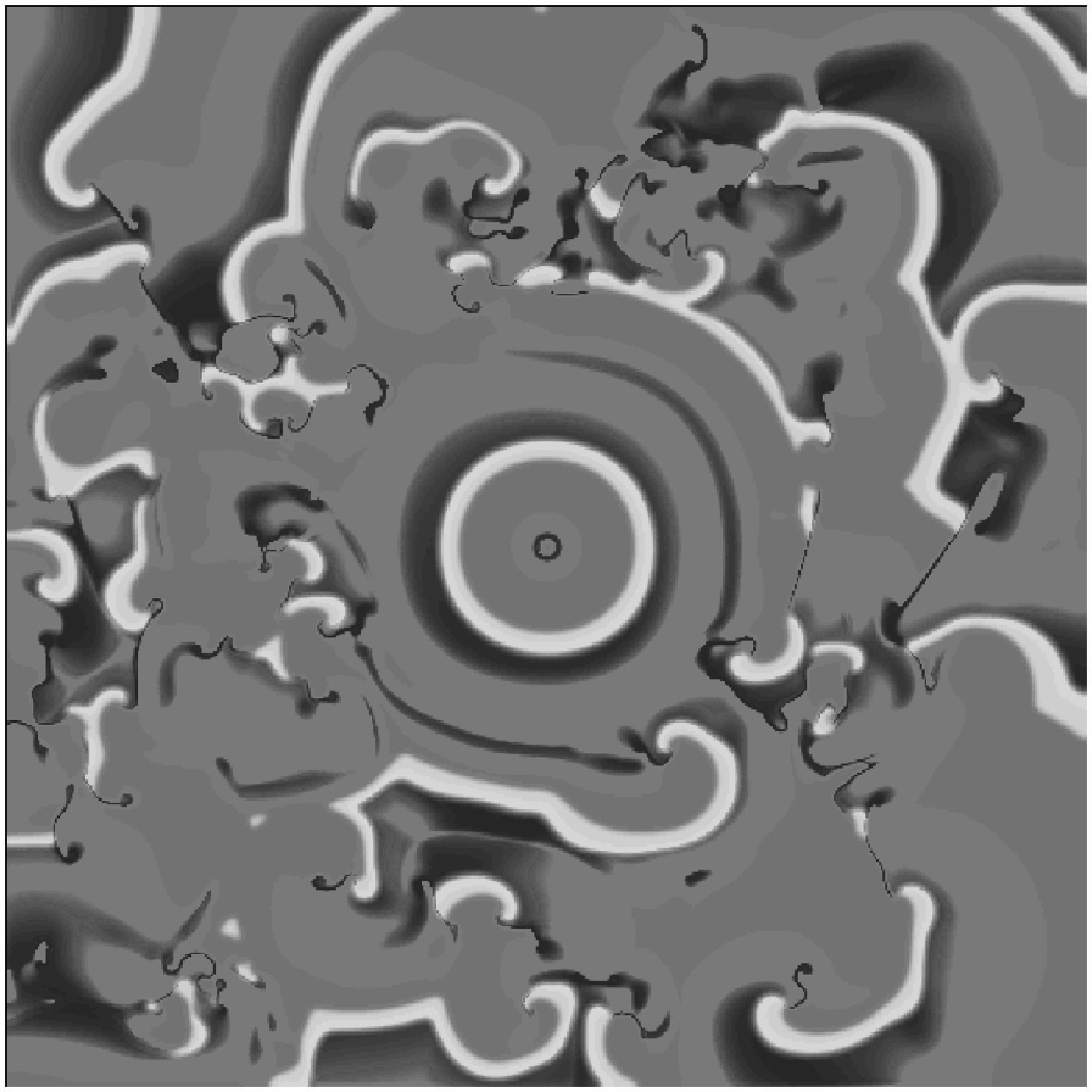}~~
\includegraphics[width=.24\textwidth]{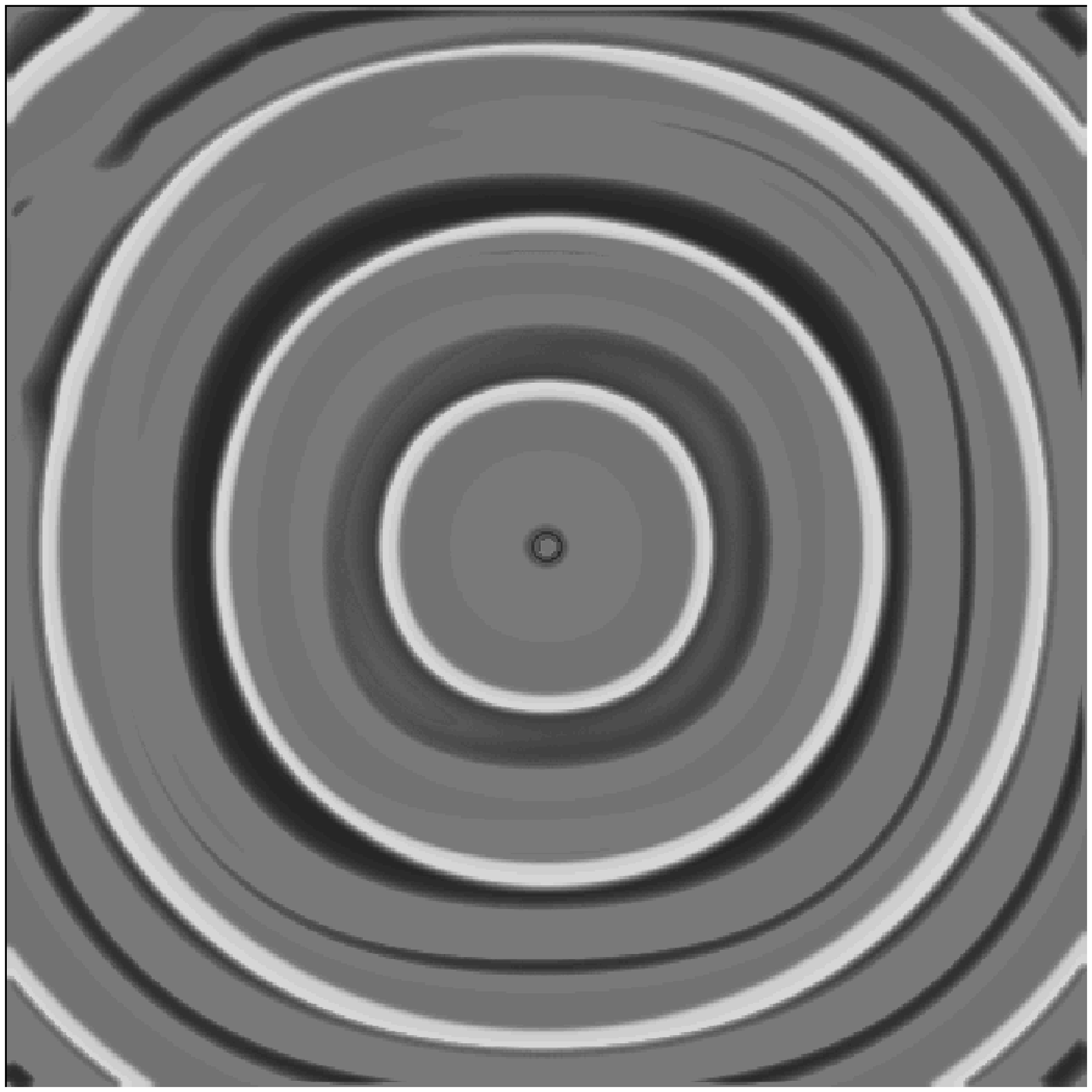}
\caption{(left) Pacing response diagram for 2D Panfilov
model ($L = 26$) showing relative performance of different waveforms.
The dash-dotted line represents a sine wave and the solid curve
represents a wave of biphasic rectangular pulses,
such that they have the same total energy. Successful control occurs
if the effective frequency lies above the broken line representing
the effective frequency of chaos ($f_c$), as seen
for 
a larger system ($L = 500$) at times $T = 1000$ (center) and 
$T = 3800$ (right) time units, where the control signal is applied 
only at the center of the simulation domain.
The excitation wavefronts are
shown in white, black marks the recovered regions ready to be excited,
while the shaded regions indicate different stages of recovery.
}
\label{ss:fig5}
\end{figure*}
We now turn to the possibility of controlling spatiotemporal chaos by 
applying control at only a small localized region of the spatially 
extended system. Virtually all the proposed local control methods 
use {\em overdrive pacing}, 
generating a series of waves with frequency
higher than any of the existing excitations in the spiral turbulent
state. As low-frequency activity is progressively invaded by faster
excitation, the waves generated by the control stimulation gradually
sweep the chaotic activity to the system boundary where they are absorbed.
Although we cannot speak of a single frequency
source in the case of chaos, the relevant timescale is that
of the spiral waves and is related to the recovery period
of the medium. 
Control is manifested as a gradually growing region in which the waves
generated by the control signal dominate, until the region expands 
to encompass the entire system. The time required to achieve termination
depends on the frequency difference between the control stimulation
and that of the chaotic activity, with control being achieved faster when
this difference is greater. 
 
Stamp {\em et al} \cite{ss:Stamp02} has looked at the possibility of using 
low-amplitude, high-frequency pacing using a series of pulses to terminate
spiral turbulence. However, using a series of pulses (having various waveform 
shapes) has met with only limited
success in suppressing spatiotemporal chaos. By contrast, a periodic
stimulation protocol \cite{ss:Zhang03} has successfully controlled chaos in
the 2D Panfilov model, as well as other models \cite{ss:note4}.
The key
mechanism underlying such control is the periodic alternation between positive 
and negative stimulation. A more general control scheme proposed in
Ref.~\cite{ss:Breuer04} uses {\em biphasic pacing}, 
i.e., applying a series
of positive and negative pulses, that shortens the recovery period around 
the region of control stimulation, and thus allows the generation of
very high-frequency waves than would have been possible using positive 
stimulation alone. A simple argument shows why a negative rectangular pulse 
decreases the recovery period for an excitable system. The stimulation
vertically displaces the $e$-nullcline and
therefore, the maximum value of $g$ that can be attained
is reduced. Consequently, the system will recover faster
from the recovery period (Fig.~\ref{ss:fig1a}, bottom).

To understand how negative stimulation affects the response behavior of 
the spatially extended system, we can use {\em pacing response diagrams}
(Fig.~\ref{ss:fig5}, left) indicating the relation between the control 
stimulation frequency $f$ 
and the effective
frequency $f_{eff}$ , measured by applying a series of pulses
at one site and then recording the number of pulses that
reach another site located at a distance without being
blocked by a region in the recovery period. Depending on the relative
 value of $f^{-1}$ and the recovery period, we observe instances of 
$n : m$
response, i.e., $m$ responses evoked by $n$ stimuli. If, for any range
of $f$, the corresponding $f_{eff}$ is significantly higher than the
effective frequency of spatiotemporal chaos, then termination of
spiral turbulence is possible. As shown in Ref.~\cite{ss:Breuer04}, there 
are indeed ranges of stimulation frequencies that
give rise to effective frequencies that dominate chaotic activity.
As a result, the periodic waves emerging from the stimulation region
gradually impose control over the regions exhibiting chaos (Fig.~\ref{ss:fig5}).
Note that,
there is a tradeoff involved here. If $f_{eff}$ 
is only slightly higher than the chaos frequency, control takes too long. 
On the other hand, if it is too high the waves suffer
conduction block at inhomogeneities produced by chaotic
activity which reduces the effective frequency, and therefore, control
fails.

Recently, another local control scheme has been proposed \cite{ss:Zhang05}
that periodically perturbs the model parameter governing the threshold.
In fact, it is the local control analog of the global control scheme proposed
by Alonso {\em et al} \cite{ss:Alonso03} discussed in 
section~\ref{ssch:sec3}.
As in the other methods discussed here, the local stimulation generates
high-frequency waves that propagate into the medium and suppress spiral
or scroll waves. Unlike the global control scheme, $b_f > > b_0$, so
that the threshold can be negative for a part of the time. This means
that the regions in resting state can become spontaneously excited, which 
allow very high-frequency waves to be generated.

\section{Discussion}
\label{ssch:sec6}
Most of the methods proposed for controlling spatiotemporal 
chaos in excitable media involve applying
perturbations either globally or over a spatially extended
system of control points covering a significant proportion
of the entire system. However, in most practical situations this may
not be a feasible option, either for issues of implementation, or because
of the high power for the control signal that such methods would need. Moreover, 
if one is using such methods in the
clinical context, e.g., terminating fibrillation, a local control scheme
has the advantage that it can be readily implemented
with existing hardware of the Implantable Cardioverter
Defibrillator (ICD). 
This is a device implanted into patients at high 
risk from fibrillation that monitors the heart
rhythm and applies electrical treatment when necessary
through electrodes placed on the heart wall. A low-energy control 
method involving ICDs should therefore
aim towards achieving control of spatiotemporal chaos by
applying small perturbations from a few local sources.

However, the problem with most local control schemes proposed so far is
that they use
very high-frequency waves to overdrive chaos. Such waves are themselves
unstable and may break during propagation, resulting in reinitiation of
spiral waves after the original chaotic activity has been terminated.
The problem is compounded by the existence of inhomogeneities in real
excitable media. Recently, Shajahan {\em et al} \cite{ss:Shajahan06} have 
found complicated dependence of spatiotemporal chaos on the presence
of non-conducting regions and other types of inhomogeneities in an
excitable system. Such inhomogeneities make the proposed local control schemes
more vulnerable, as it is known that high-frequency pacing interacting
with, e.g., non-conducting obstacles, results in wave breaks and
subsequent genesis of spatiotemporal chaos \cite{ss:Panfilov93a}.

The search is still on for a control algorithm for terminating spatiotemporal 
chaos in excitable media, that can be implemented using low power, or, that
need be applied in only a small, local region of the system, and which 
will yet be robust, capable of terminating spiral turbulence without
the control stimulation itself breaking up subsequently. The payoffs for coming
up with such a method are enormous, as the potential benefits include
an efficient device for cardiac defibrillation. 

\vspace{0.5cm}
\noindent
{\bf Acknowledgements:} We would like to thank colleagues and 
collaborators with whom some of the work described above has been carried
out, especially, Rahul Pandit, Ashwin Pande, Avishek Sen, T. K. Shajahan,
David J. Christini, Ken M. Stein, Johannes Breuer and Antina Ghose.
We thank IFCPAR and IMSc Complex Systems Project (X Plan) for support.


\begin{thebibliography}{999}
\bibitem{ss:note1} {Note that, unlike waves in ordinary
diffusive media which dissipate as they propagate further, excitation waves
are self-regenerating.}.
\end{thebibliography}
\end{document}